\documentclass[12pt,letter]{article}

\usepackage{graphicx, epsfig, color, subfigure, amssymb,ulem}
\def\eslt{\not\!\!{E_T}}
\def\eslt{E_T^{\rm miss}}
\def\to{\rightarrow}

\def\bi{\begin{itemize}}
\def\ei{\end{itemize}}

\def\tu{\tilde u}

\def\td{\tilde d}

\def\tg{\tilde g}
\def\tq{\tilde q}

\def\tell{\tilde\ell}
\def\tq{\tilde q}

\def\tz{\widetilde Z}

\def\alt{\lesssim}
\def\agt{\gtrsim}
\def\mt2{m_{T2}}
\def\be{\begin{equation}}  
\def\ee{\end{equation}}  
\def\msq{m_{\tilde q}}
\def\mg{m_{\tilde g}}

\def\pT{\mathbf{p_T}}

\newcommand\prd[3]{{\it Phys.\ Rev.\ }{\bf D #1} (#2) #3}

\newcommand\prl[3]{{\it Phys.\ Rev.\ Lett.\ }{\bf #1} (#2) #3}
\newcommand\plb[3]{{\it Phys.\ Lett.\ }{\bf B #1} (#2) #3}
\newcommand\jhep[3]{{\it J. High Energy Phys.\ }{\bf #1} (#2) #3}

\newcommand\npb[3]{{\it Nucl.\ Phys.\ }{\bf B #1} (#2) #3}

\newcommand{\hepph}[1]{hep-ph/#1}

\newcommand{\bea}{\begin{eqnarray}}
\newcommand{\eea}{\end{eqnarray}}

\newcommand\ck[1]{\textcolor{blue}{#1}}
\begin{document}
\begin{titlepage}
\begin{flushright}
%BRA-HEP/000001\\
MADPH-11-1569

UH-511-1158-11
\end{flushright}

\vspace{0.5cm}
\begin{center}
{\Large \bf 
Determining the squark mass at the LHC
}\\ 
\vspace{1.2cm} \renewcommand{\thefootnote}{\fnsymbol{footnote}}
{\large Vernon Barger $^1$\footnote[1]{Email: barger@physics.wisc.edu },
Yu Gao $^2$\footnote[2]{Email: ygao3@uoregon.edu},
Andre Lessa $^3$\footnote[3]{Email: lessa@nhn.ou.edu },
Xerxes Tata $^4$\footnote[4]{Email: tata@phys.hawaii.edu } \\
\vspace{1.2cm} \renewcommand{\thefootnote}{\arabic{footnote}}}
{\it 
1. Physics Dept. University of Wisconsin, Madison, WI 53706, USA\\
2. Dept. of Physics, University of Oregon, Eugene, OR 97403, USA\\
3. Dept. of Physics and Astronomy, University of Oklahoma, Norman, OK 73019 USA\\
4. Dept. of Physics and Astronomy,
University of Hawaii, Honolulu, HI 96822, USA\\
}

\end{center}

\vspace{0.5cm}
\begin{abstract}
\noindent 
We propose a new way to determine the squark mass based on the shape of
di-jet invariant mass distribution of supersymmetry (SUSY) di-jet events
at the Large Hadron Collider (LHC).  Our algorithm, which is based on
event kinematics, requires that the branching ratio $B(\tq \to q\tz_1)$
is substantial for at least some types of squarks, and that
$m_{\tz_1}^2/m_{\tq}^2 \ll 1$.  We select di-jet events with no isolated
leptons, and impose cuts on the total jet transverse energy,
$E_T^{tot}=E_T(j_1)+E_T(j_2)$, on $\alpha = E_T(j_2)/m_{jj}$, and on the
azimuthal angle between the two jets to reduce SM backgrounds. The shape
of the resulting di-jet mass distribution depends sensitively on the
squark mass, especially if the integrated luminosity is sufficient to
allow a hard enough cut on $E_T^{tot}$ and yet leave a large enough
signal to obtain the $m_{jj}$ distribution.  We simulate the signal and
Standard Model (SM) backgrounds for 100 fb$^{-1}$ integrated luminosity
at 14 TeV requiring $E_T^{tot}> 700$~GeV. We show that it should be
possible to extract $m_{\tq}$ to within about 3\% at 95\% CL --- similar to
the precision obtained using $m_{T2}$ --- from the di-jet mass distribution
if $m_{\tq} \sim 650$~GeV, or to
within $\sim 5$\% if $m_{\tq}\sim 1$~TeV.

\vspace{0.8cm}
\noindent PACS numbers: 14.80.Ly, 12.60.Jv, 11.30.Pb, 13.85.Rm
\label{bloc:abstract}
\end{abstract}

\end{titlepage}

\section{Introduction}
\label{sec:intro}

Supersymmetry (SUSY) is one of the best-motivated extensions of the
Standard Model (SM)~\cite{rev}. There have been numerous studies of
strategies by which superpartners may be discovered at the LHC. The
SUSY reach of the LHC is usually expressed in terms of
the masses of coloured gluinos and squarks, expected to be the most
copiously produced sparticles~\cite{reach,atlas,cms}.  More recently,
the focus has shifted to how well sparticle properties can be
determined at the LHC, with the most attention being paid to sparticle
masses, the underlying motivation being that knowledge of the
sparticle spectrum will lead us to the mechanism by which SM
superpartners acquire their masses. The problem, of course, is that it
is not possible to construct mass peaks because every SUSY event (in
$R$-parity conserving SUSY models that we focus upon in this paper)
includes two undetected neutralinos ($\tilde{Z}_1$), which we take to be the lightest
supersymmetric particle (LSP).

The variable $M_{\rm eff}\equiv \sum_{i=1}^4 E_T(j_i)+|\eslt|$
constructed from transverse energy ($E_T$) provides a rough measure of
the mass of the lighter of gluinos/squarks~\cite{hinchliffe,atlas}. The
end point of the same flavour, opposite-sign dilepton mass spectrum,
first mentioned in ~Ref.\cite{dilep}, yields a precise measurement of
the {\it mass difference}, $m_{\tz_2}-m_{\tz_1}$, between
neutralinos. Hinchliffe, Paige and collaborators~\cite{hinchliffe}
pioneered systematic studies of the information that may be gleaned from
end-points of a variety of mass distributions in LHC event samples with
cuts to select out SUSY events over SM backgrounds, and showed that
(modulo some discrete ambiguities) in favourable cases with two-body
decay cascades, it is sometimes possible to also construct the masses,
rather than just mass differences: in this connection, see also
Ref.~\cite{gunion}.  Since then, other techniques have been suggested to
obtain sparticle masses. These include the use of the $m_{T2}$
variable~\cite{mt2}, its cousins $m_{T\rm{Gen}}$ and related variables
~\cite{mtgen}, as well as their generalization to asymmetric
decays~\cite{asymm}, the so-called matrix element method~\cite{matrix},
the presence of kinematic cusps in distributions~\cite{cusps}, or
through multi-lepton channels~\cite{bib:lep}.

In this paper, we propose a new way of measuring masses for the case of
event topologies with one step cascade decays of the type:
\begin{equation}
P^{(1)} + P^{(2)} \rightarrow q^{(1)} + D^{(1)} + q^{(2)} + D^{(2)} 
\label{eq:cdec}
\end{equation}
where $P^{(i)}$ are the parent particles (squarks in our analysis)
initially produced in the hard scattering, $D^{(i)}$s are the invisible
(lightest SUSY particles, LSP) daughters of $P^{(i)}$, while each
$q^{(i)}$ manifests itself as a jet in the detector. None of the
existing mass measurement methods allow for a separate determination of
both the parent and daughter masses from just the process
(\ref{eq:cdec}). For instance, the much-studied
$m_{T2}$ method yields a measure of $(m_P^2-m_D^2)/2m_P$ which,
only if $m_D^2/m_P^2 \ll 1 $, provides a
measurement of the parent mass \cite{choi}. If more complicated decay chains
\cite{Mt2sub} or several production processes \cite{choi} are
accessible, it may be possible to extract the individual parent and
daughter masses. The method presented here offers an alternative way of
determining $m_P$ (also in the regime $m_D^2/m_P^2 \ll 1 $) from the
dijet invariant mass distribution,  and relies on
completely distinct kinematical features from the standard $m_{T2}$
procedure.

In our analysis, we assume that the daughter LSP is a bino-like
neutralino $\tilde{Z}_1$, while the parent is a right squark, which
dominantly decay via $\tq_R\to q\tz_1$.\footnote{If instead, the LSP is
a wino-like neutralino, $\tq_L$ would dominantly decay to charged or
neutral winos. Since the SM daughters of the chargino decays would be
very soft, the analysis we describe for $\tq_R$ would then apply to
$\tq_L$.} Thus our signal from right squark pair production is exactly
two hard jets, no isolated leptons or photons, together with missing
transverse energy ($\eslt$).  It has been pointed out~\cite{rts}, and
since confirmed \cite{lessa,LHCalpha} that SM backgrounds to this signal
can be controlled by requiring acollinear jets, even without the use of
$\eslt$. If gluinos are marginally heavier than squarks, they decay to
squarks and an undetectably soft jet, so that $\tg\tq$ production then
adds to the squark dijet signal. If gluinos are light enough so that
squarks dominantly decay to gluinos, we do not have the dijet plus
$\eslt$ signal, and our analysis does not apply.  
Aside
from our assumption about the mass ordering between gluinos and squarks
and $m_{\tz_1} \ll m_{\tq}$, we endeavor to leave our analysis as
model-independent as possible, and at the same time realistic in that we
include all complications due to contamination from other SUSY processes
and SM background. We will see below that in addition to yielding
$m_{\tq}$ our analysis may also serve to approximately constrain $m_{\tg}$.

\section{The acollinear dijet signal}
\label{sec:acoll}

The cross section for pair production of $\tq_R$ is determined by SUSY
QCD just in terms of $m_{\tg}$ and $m_{\tq_R}$, independent of the
details of any model. For the mass ordering of interest to us, the decay
$\tq_R\to q\tg$ is kinematically suppressed, so that right squarks
dominantly decay to the (bino-like) LSP. SUSY contamination to the
acollinear dijet signal is, of course, model-dependent. For
definiteness, we assume gaugino mass unification at the GUT scale,
take $\mu = m_{\tg}$, $\tan\beta=10$, and use the scalar mass unification
condition $m_{\tq}^2\simeq m_{\tell}^2 + 0.7 m_{\tg}^2$ as a guide to
the slepton mass. These choices have little effect on the right squark
signal other than restricting the gluino to be not much heavier than the
squark, since otherwise the slepton becomes too light.

We use ISAJET v7.78~\cite{isajet} for our simulation of the SUSY
signal at the LHC. We assume a toy calorimeter with cell size
$\Delta\eta\times\Delta\phi=0.05\times 0.05$, extending to $|\eta|=4$. 
Here $\eta,\phi$ denote the jet pseudo-rapidity and azimuthal angle.
The hadronic calorimetry (HCAL) energy resolution is taken to be
$100\%/\sqrt{E_T}\oplus 5\%$, where the two terms are combined in
quadrature. The electromagnetic calorimetry (ECAL) energy resolution
is assumed to be $5\%/\sqrt{E_T}\oplus 0.55\%$. We use the cone type ISAJET
jet finding algorithm, with a cone size $\Delta
R\equiv\sqrt{\Delta\eta^2+\Delta\phi^2}\leq0.4$ to group HCAL energy
depositions into jets.  Jets are then defined to be hadronic clusters
with $E_T(jet)>50$ GeV.  Leptons with transverse momentum $p_T(l)>5 $ GeV are defined to
be isolated if the visible activity within a cone of $\Delta R<0.2$
about the lepton direction satisfies $\Sigma E_T^{cells}<5$ GeV.

In the analysis of our signal, we require exactly two jets with, 
\be
\begin{array}{cl}
1.&\hspace{2mm} E_T(j)\ge \textrm{50 GeV}, \\
2.&\hspace{2mm} \textrm{ azimuthal angle separation } \Delta\phi_{jj} \le
1.5,\\   
3.&\hspace{2mm} E_T^{tot}\equiv E_T(j_1)+E_T(j_2)> 700 \textrm{~GeV, and} \\
4.&\hspace{2mm} \alpha \equiv E_T(j_2)/m_{jj} \ge \textrm{0.5.}
\end{array}
\label{eq:cuts}
\ee where $j_1$, $j_2$ are ordered by jet $E_T$ ($E_T(j_1)>E_T(j_2)$)
and $m_{jj}$ is the invariant mass of the two jets.  We veto events with
isolated leptons or a third jet with $E_T(j)\ge 50$~GeV.  These cuts and
veto criteria have been shown to very effectively remove the QCD dijet
background~\cite{lessa,LHCalpha}.  The SM background is then dominated
by $(Z \to \nu\bar{\nu})+jj$ events~\cite{lessa,LHCalpha}, which has a
cross section $\sigma_{Zjj}= $~6~fb, with an additional contribution
of $\sigma_{Wjj}=1.8$~fb from $W \to \ell\nu$ events where the lepton is
not identified. 

\begin{figure}
\begin{center}
  \includegraphics[scale=0.7]{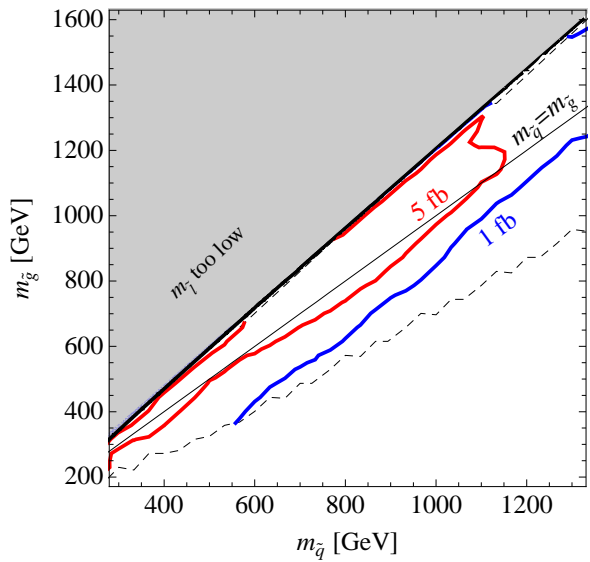}
\end{center}
\caption{SUSY signal cross section contours for the acollinear dijet
 signal from supersymmetry at a 14~TeV $pp$ collider, after  cuts 
 listed in Eq.~(\ref{eq:cuts}). 
 The dashed line shows the boundary of sampled area taken approximately
 to be $m_{\tg}\simeq 0.7m_{\tq}$.  
 The upper left part of the plane is cut off because, for our choice of
 slepton masses, this region is forbidden either because the slepton is
 lighter than $\tz_1$ or $m_{\tilde{l}}$ drops below $100$~GeV. The
 cross section is largest just above $\msq=\mg$ where it can
 easily exceed 10 fb. }
\label{fig:csec}
\end{figure}

Contours of the signal cross section at a 14~TeV $pp$ collider after the
cuts listed above are shown in the $m_{\tq}-m_{\tg}$ plane in
Fig.~\ref{fig:csec}. We cut the contours off at large values of
$m_{\tg}$ where the slepton either becomes lighter than $\tz_1$, or when
it violates the LEP lower mass limit on its mass. The cross section
drops off when $m_{\tg}$ falls significantly below $m_{\tq}$ because
squark decays to gluinos begin to eat into the branching ratio $B(\tq
\to q\tz_1)$. The region where we terminate the 1~fb contour has
recently been excluded at 95\% CL by the CMS and ATLAS experiments
at the LHC \cite{cms_bnd}.

When the gluino is heavier than the squark the
$\tilde{g}\rightarrow\tilde{q}+\bar{q}$ decay channel opens up, and
$\tg\tq$ 
events can pass the $N_j = 2$ cut if the $\bar{q}$ forms a very soft jet,
below the jet $E_T$ threshold. This  roughly doubles the signal just above
the diagonal $m_{\tg}=m_{\tq}$ line. However, this contribution falls off
with increasing $m_{\tg}$ (and fixed $m_{\tq}$) because the additional
jet becomes more readily identified in the detector.
If instead $m_{\tg} < m_{\tq}$, the
$\tilde{q}\rightarrow\tilde{g}+q$ decay channel takes over quickly as
the mass difference increases. This decay channel leads to multiple jets
and the signal drops below 1 fb at $\mg\sim \frac{4}{5} \msq$.  The
signal is also suppressed at low $(m_{\tq},m_{\tg})$ due to the
$E_T^{tot}$ cut. Nevertheless, we see that there is a sizable region
where there are several hundred events in the SUSY dijet 
channel for an integrated luminosity of
100 fb$^{-1}$.

\section{Standard Model backgrounds} \label{sec:back}

The dominant SM background to the SUSY signal in the acollinear dijet
channel comes from the (Z$\rightarrow \nu\bar{\nu}) + 2j$  events.
Because the properties of jets in $Z+2j$ events at the LHC cannot depend
on how the $Z$ decays, this background can be directly obtained
\footnote{Here, we are assuming that SUSY does not significantly contribute to $Zjj$
  events.}  by
scaling LHC data on
$Z\rightarrow e^+e^-+\mu^+\mu^-$+2 jet events by a 
factor,
\be
\xi=\frac{BR(Z\rightarrow \nu\bar{\nu})}{BR(Z\rightarrow
e^+e^-/\mu^+\mu^-)}\simeq 3.0.
\ee
 The lower rate in the
(Z$\rightarrow \ell\bar{\ell}$)+$jj$ channel, however, implies
that the fluctuations in the inferred background will be larger
by $\sqrt{\xi}$ relative to the corresponding fluctuations in the 
$Z (\to \nu\bar{\nu}) + 2j$ sample.
In our analysis we have used the Monte Carlo
event generator AlpGen~\cite{Mangano:2002ea} to simulate the SM
(Z$\rightarrow \nu\bar{\nu}$)+$jj$ background, and then  scaled the
uncertainty in each $m_{jj}$ bin by $\sqrt{\xi}$ to correctly
simulate  the
corresponding fluctuations when this background is obtained from 
the $Z\to\ell{\bar{\ell}}+jj$ data. 

In addition, as already mentioned, $W + jj$ events make a
subdominant but sizeable contribution to the background. Occasionally the
charged lepton (most likely a hadronically decaying $\tau$) from a high
$p_T$ W decay becomes buried in the hadronic jets and/or is undetected
because it is either too soft or in uninstrumented regions of the
detector, while the neutrino is hard enough to reduce the angular
separation of the two jets. The relatively small $W+jj$ background
compared to the leading $Z+jj$ one reduces the relevance of $W +jj$
contamination. While the subtraction of the $Wjj$ background is not as
simple as for the background from $Zjj$ production, we may expect that
the $W+jj$ signal can be measured in a control region (such as
the dijet + 1 lepton channel) and then extrapolated to the signal region
(dijet + 0 leptons).\footnote{This could be complicated by the fact that
SUSY events may also make a contribution to the jets plus lepton
channel, but typically, one would expect that the signal jet
multiplicity is larger.}  In our analysis, just as for the $Zjj$
background, we assume that we can use AlpGen \cite{Mangano:2002ea} to
simulate and subsequently subtract the $Wjj$ background from the total
di-jet sample, but include the corresponding statistical uncertainty in
the evaluation of $\chi^2$ when performing our fits to extract sparticle
masses.

\begin{figure}
\begin{center}
\includegraphics[scale=0.7]{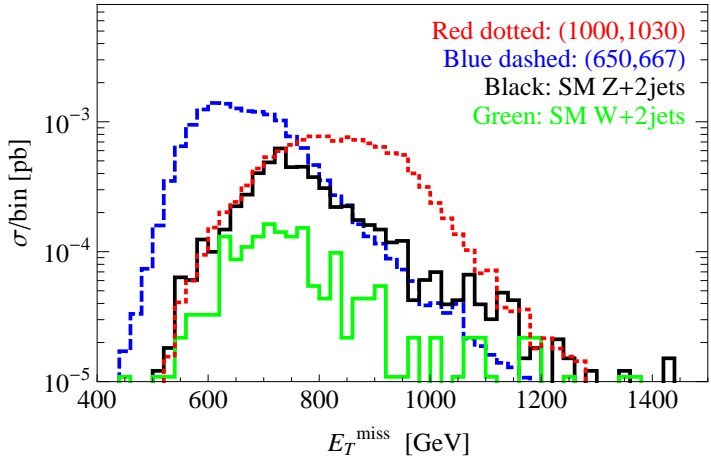}
\end{center}
\caption{The $\eslt$ distribution for the leading SM backgrounds from $Z+jj$
and $W+jj$ production, together with the same distributions for
the signal points $(m_{\tq},m_{\tg})=(650,667)$~GeV and
$(1000,1030)$~GeV at a 14 TeV \textit{pp} collider after the cuts in
Eq.~(\ref{eq:cuts}). The $Z+jj$ background shown here
is the SM prediction for invisible $Z$ decays and does not show
the enhanced fluctuations that would result if this background
were extracted from the $(Z\to
e^+e^-/\mu^+\mu^-)+jj$ data as discussed in the text.}
\label{fig:bg}
\end{figure}

The reader may well wonder whether it is further possible to enhance the
SUSY signal relative to background via a requirement on $\eslt$.  With
this in mind, we show the $\eslt$ distribution of SM $Z/W$ + 2 jet
backgrounds in Fig.~\ref{fig:bg}, along with the
corresponding distribution for the two SUSY test cases that we introduce
below for our study of how well the squark mass may be extracted at the
LHC. The total SM background is 7.8 fb after the cuts
in Eq.~(\ref{eq:cuts}) to be compared with the signal of 16~fb (11~fb) for
$m_{\tq}=650$~GeV (1~TeV).  Note that the angular and $E_T^{tot}$ cuts
preclude very low values of $\eslt$. As the signal $\eslt$ distribution
moves to higher values with increasing $\msq$, it may appear tempting to
impose an additional $\eslt$ cut at 750 GeV to further enhance the
signal events relative to the background. We have checked, however, that
the loss in signal statistics largely negates the benefit of a lower
background; the $\eslt$ cut does not yield a significant improvement to the
determination of $\msq$. 

\section{Squark mass determination}
\subsection{Methodology}\label{sec:sqmass}

\begin{figure}

\begin{center}
  \subfigure[]{\includegraphics[scale=0.65]{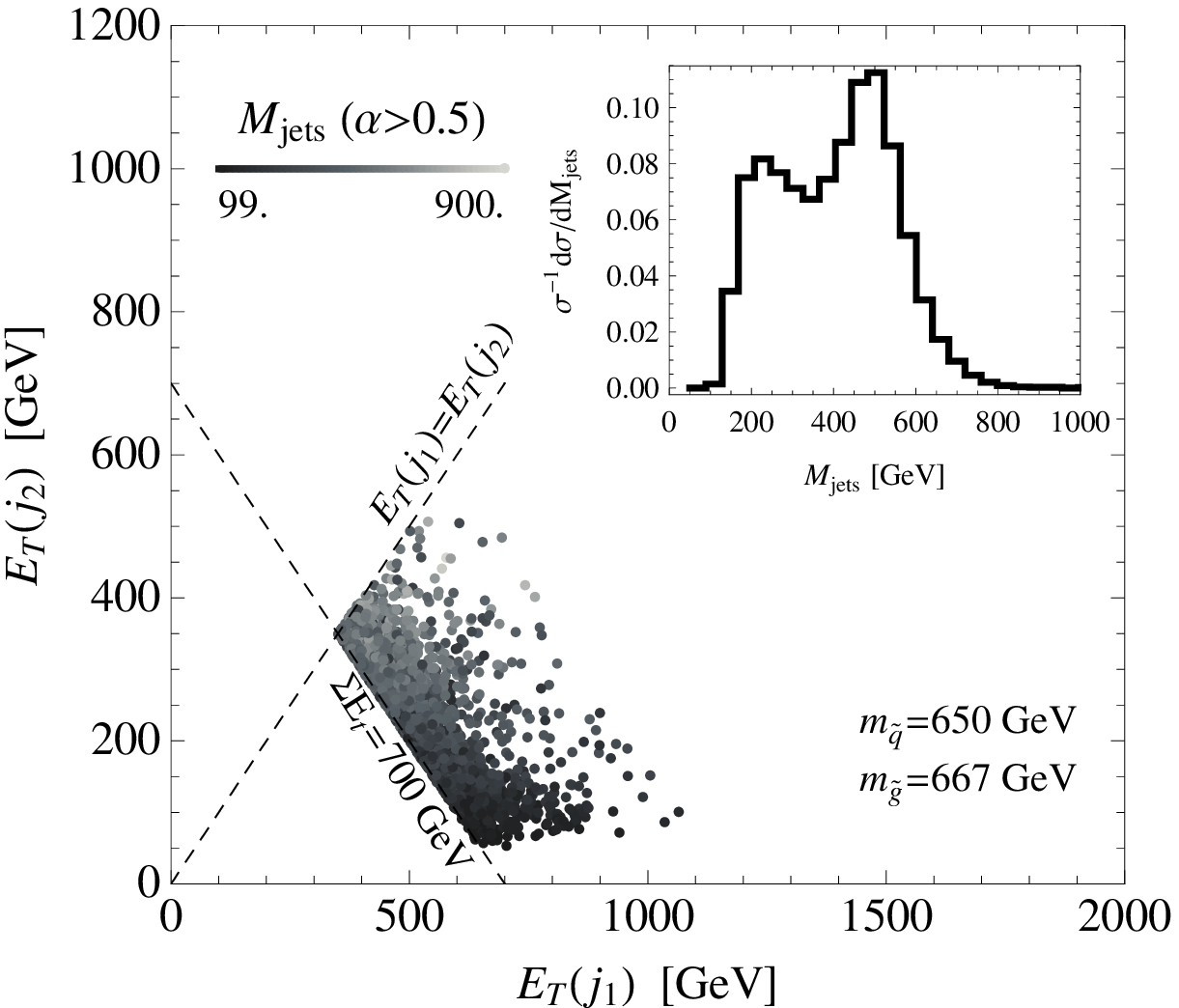}}
  \subfigure[]{\includegraphics[scale=0.65]{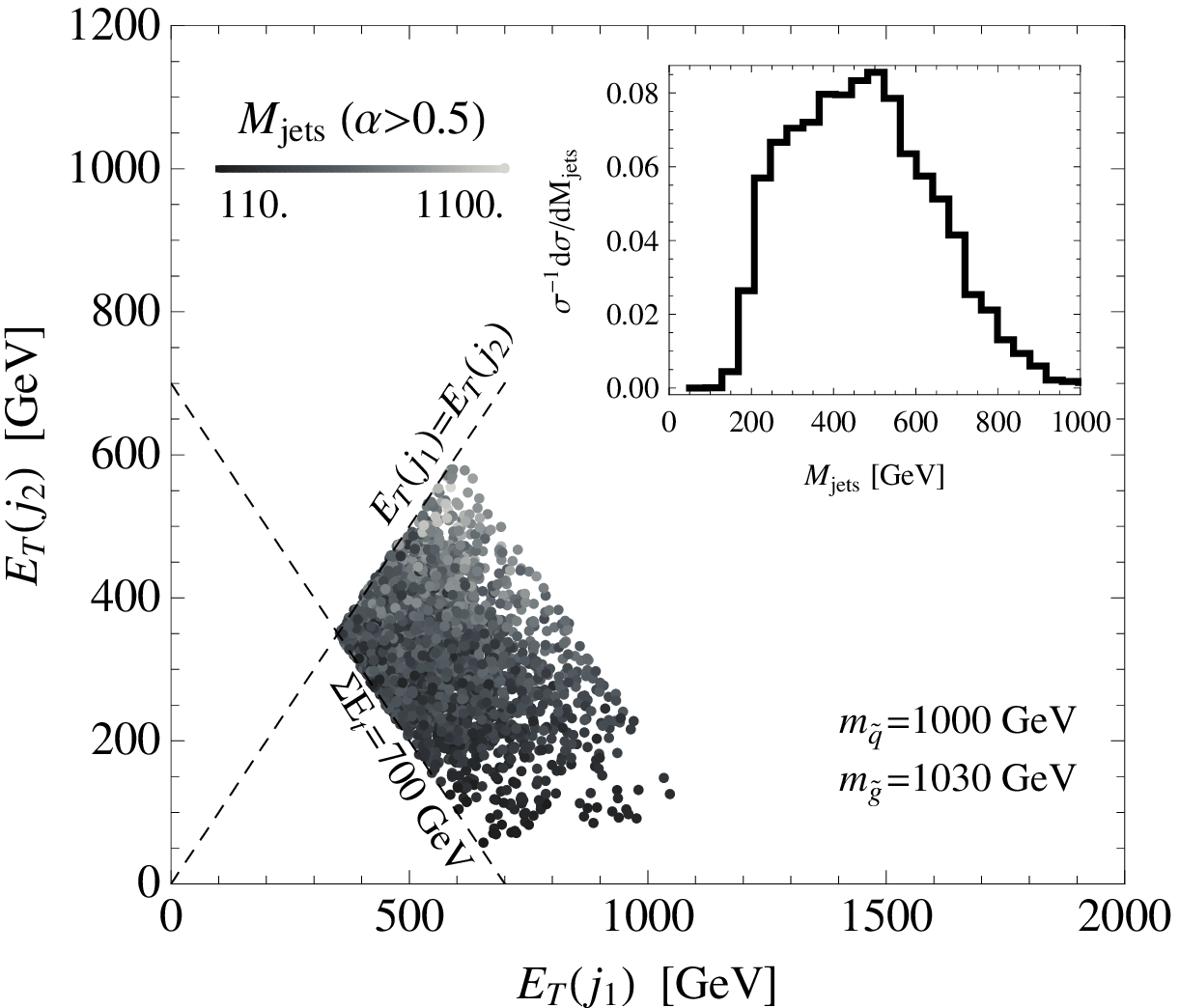}}
\end{center}
\caption{A scatter plot of the jet transverse energies for dijet events
 from supersymmetry at a 14~TeV $pp$ collider for
 Frame (a)~$m_{\tilde{q}}$ = 650~GeV, and
 Frame (b)~$m_{\tilde{q}}$ = 1 TeV after the ($\alpha, \Delta\phi_{jj}, E_T^{tot}$) 
 cuts described in the
 text. The gluino mass is chosen slightly higher than the squark mass to
 enhance event rates by including gluino production channels.
 The invariant mass of the dijets is shown by the gray-scale on
 the dots. Notice that the region with the largest $E_T(j_1)$ at the
 lower part of the plot is populated by events with low invariant mass,
 while events with high values of $E_T(j_2)$ and of $m_{jj}$
 preferentially populate the upper part of the plot. This leads to the
 shoulders in the $m_{jj}$ distribution for the 650 GeV squark case
 shown in the inset frame; this feature, absent for the 1~TeV
 squark case, would appear if we require
 $E_T^{tot}>1$~TeV.}

\label{fig:scatter}
\end{figure}

To facilitate the discussion of our strategy for squark mass
determination, we show a scatter plot of the two jet energies in
Fig.~\ref{fig:scatter} for ({\it a})~$m_{\tq}=650$~GeV, and ({\it
b})~$m_{\tq}=1$~TeV. Although we include {\it all}  SUSY events in this
sample, about 40\% ($>90$\%) of the events come from squark pair
production in frame ({\it a}) where the gluino-squark mass gap, $\Delta
M$, is just 17~GeV, (in frame ({\it b}) with $\Delta M=30$~GeV) with the
bulk of the remaining events arising from $\tg\tq_R$ pair production. In
the former case, although squark production seems to be subdominant, we
should remember that the gluino decays into a squark and a (very soft)
jet, so that these contaminating $\tq\tg$ events have very similar
kinematics as squark-pair events. Hence $\tq\tg$ events do not interfere with our
strategy to extract the squark mass and, in fact, help in that they
increase the signal sample that we can use.\footnote{If the
gluino-squark mass gap becomes substantial, the additional jet fom
gluino decay frequently fails the third jet veto, suppressing the gluino
signal. In case $\mg\gg\msq$ the gluino production is suppressed and
$\tq\rightarrow q\tilde{Z}_1$ remains the only viable signal.  }

The gray scale on the scatter plot shows how the events are distributed
by invariant mass. We see that the bulk of events with low invariant
dijet mass $m_{jj}$ have small values of $E_T(j_2)$ and (because of the
$E_T^{tot}$ cut) correspondingly large values of $E_T(j_1)$. This is not
surprising since (except when both jets are very forward) our jet cone
algorithm generally requires the jets to have a spatial angular
separation between them, and precludes the smallest invariant masses if
$E_T(j_2)$ is also sizeable. Despite the fact that $j_2$ is soft, these
events readily satisfy the $\alpha$ requirement because $m_{jj}$ is also
small.  Relative to these events with the lowest values of $m_{jj}$,
events with somewhat larger dijet masses are possible if {\it either}
$E_T(j_2)$ or the space angle between the jets is increased.  If
$m_{jj}$ becomes larger because $E_T(j_2)$ is increased, the $\alpha$
cut is still satisfied.  However, events with increased values of
$m_{jj}$ due to a larger angular separation but where $E_T(j_2)$ remains
small have smaller values of $\alpha$, and so a reduced efficiency for
passing the $\alpha>0.5$ cut. We emphasize that {\it without the
$E_T^{tot}$ cut} these ``intermediate $m_{jj}$ events'' would also be
obtained for smaller values of $E_T(j_1)$ and correspondingly larger
values of $E_T(j_2)$ that would readily satisfy $\alpha > 0.5$.
Finally, events with large values of $E_T(j_2)$ close to the
$E_T(j_1)=E_T(j_2)$ line naturally tend to have high values of dijet
masses (as evidenced by the lack of dark points in this region of the
scatter plot) as well as of $\alpha$.  Of course, events with exactly
back-to-back jets are {\it always} eliminated by the $\alpha$ cut (and
also the $\Delta\phi$ cut), which is why this cut is an efficient veto
against QCD dijet events.

To recap, these qualitative arguments suggest that, compared to events
with very small or relatively large values of $m_{jj}$, events with
intermediate values of dijet invariant mass are less likely to satisfy a
hard cut on $E_T^{tot}$ simultaneously with the requirement $\alpha >
0.5$.  The dip in the di-jet invariant mass distribution near
$m_{jj}\sim 350$~GeV in Fig.~\ref{fig:scatter}({\it a}) is a
manifestation of exactly this feature. This deficit of intermediate
events, as we will see below, which plays an important role in the
extraction of $m_{\tq}$ from the di-jet data, does not show up in the
1~TeV case in frame ({\it b}). To better understand this, we have
analysed the kinematics of di-jet events from squark decay in the
Appendix in more detail. In the approximation that the hard jet $j_1$ is
in the direction of its squark parent, we show that with $E_T^{tot}>
2E_0 \equiv (m_{\tq}^2-m_{\tz1}^2)/m_{\tq}$, high mass events with large
opening angle between the jets, as well as low mass events with a small
opening angle between the jets readily satisfy that $\alpha > 0.5$
requirement. We find, however, that intermediate mass events where the
second jet is roughly perpendicular to the hard jet (in the squark CM
frame) {\it can never} satisfy the $\alpha$ cut if $E_T^{tot}>
2.6E_0$. Although our analysis in the Appendix neglects events from
$\tg\tq$ production for which the kinematics is slightly different
because of the soft jet from gluino decay, and also any QCD radiation,
it nevertheless confirms the general features that we inferred from our
qualitative arguments of this Section. More importantly, it fixes the
lower limit on $E_T^{tot}$ in order for the dip to become apparent in
the $m_{jj}$ distribution to be $\simeq 2E_0 \simeq m_{\tq}$ if $m_{\tz_1}\ll
m_{\tq}$. We now understand why we do not see a corresponding dip in
Fig.~\ref{fig:scatter} ({\it b}) --- the $E_T^{tot}> 700$~GeV cut is not
hard enough, and the would-be dip region is populated by events with
not-so-large values of $E_T(j_1)$, and concomitantly larger values of
$E_T(j_2)$ that then pass the $\alpha> 0.5$ cut. Indeed, we have
verified that by hardening the $E_T^{tot}$ cut to 1~TeV we recover the
dip also in this case. We have not shown this because the event rate then
becomes too small; {\it i.e.} despite the appearance of the dip the
reduced event rate does not lead to any improvement in the determination
of $m_{\tq}$, at least for an integrated luminosity of 100~fb$^{-1}$.
For this reason, we leave the $E_T^{tot}$ cut at 700~GeV throughout this
analysis.

We fit the synthetic $m_{jj}$ data for the cuts in (\ref{eq:cuts}) to
theoretical templates for a grid of $(m_{\tq}, m_{\tg})$ in order to
extract the squark mass. We use an integrated luminosity of at least
1000~fb$^{-1}$ to generate these templates.  We will see below that the
squark mass can be extracted with greater precision if the integrated
luminosity is sufficient for the implementation of a large enough
$E_T^{tot}$ cut so that the dip structure in the di-jet mass
distribution (which is very sensitive to $m_{\tq}$) is clearly evident.

\subsection{Results}
\label{sec:massanal}

We now discuss the precision with which
the squark mass can be extracted via fits to the di-jet mass
distribution. We use the $m_{\tq}=650$~GeV and $m_{\tq}=1000$~GeV cases
introduced above as illustrative examples.
In order to extract the squark mass we fit the {\it shape} of
the signal $m_{jj}$ distribution --- obtained after statistical
subtraction of the backgrounds from the signal plus background sample 
as explained in Sec.~\ref{sec:back} --- to
the corresponding distributions obtained using templates that have been
independently generated for a grid of values of $(m_{\tq}, m_{\tg})$.
Given that a signal has been detected in the di-jet sample,
we define $\chi^2$ for the normalized signal spectrum $\phi^0\equiv
\frac{1}{N_{0}}\frac{dN_0}{d m_{jj}}$ as,

\be
\chi^2(\phi)=\sum_{i}\frac{(\phi_i
-\phi^0_i)^2}{\delta_{i,\phi^0}^2+\delta^2_{i,Zjj}+\delta^2_{i,Wjj}}\;,
\label{eq:chi2}
\ee where $i$ sums over all $m_{jj}$ bins, each one 30 GeV wide, and
bins with less than 5 signal events are dropped. Here, $\phi_i^0$
correspond to the ``data'' for 100~fb$^{-1}$ in the $i$th bin, while
$\phi_i$ is the corresponding expectation for the value of $(m_{\tq},
m_{\tg})$ from the template obtained with an integrated luminosity
exceeding 1000~fb$^{-1}$.
The uncertainty $\delta_i\equiv
\sqrt{N_i^{stat}}/N_0^{tot}$ gives the statistical error in the shape of
signal spectrum after background subtraction. In this analysis we do not
include systematics. We  minimize $\chi^2(\phi)$ over a grid in
$(m_{\tq},m_{\tg})$ values, with its minimum being the best fit for
$(m_{\tq},m_{\tg})$, and as usual map out contours of constant
$\Delta\chi^2$ in the $m_{\tq}$-$m_{\tg}$ plane. 
As mentioned in Sec.~\ref{sec:acoll},
gluino events only contribute to the signal if $m_{\tg}\sim
m_{\tq}$. The gluino contribution improves the signal statistics but also has
a small effect in the $m_{jj}$ shape, and so affects the sparticle mass
extraction as will be discussed shortly.

We note that Eq.~(\ref{eq:chi2}) assumes perfect evaluation of the
theoretical spectrum $\phi_i$. In practice, although we use an order of
magnitude larger integrated luminosity for the calculation of the
expectation from the templates, significant fluctuations remain
distorting and even fragmenting the $1\sigma$ (68\% confidence level
(CL)) contour. For this reason, we only show results for the
extraction of sparticle masses at the $2\sigma$ and $3\sigma$ contours in the
following.

\begin{figure}
\begin{center}
\subfigure[\ck{}]{\includegraphics[scale=0.72]{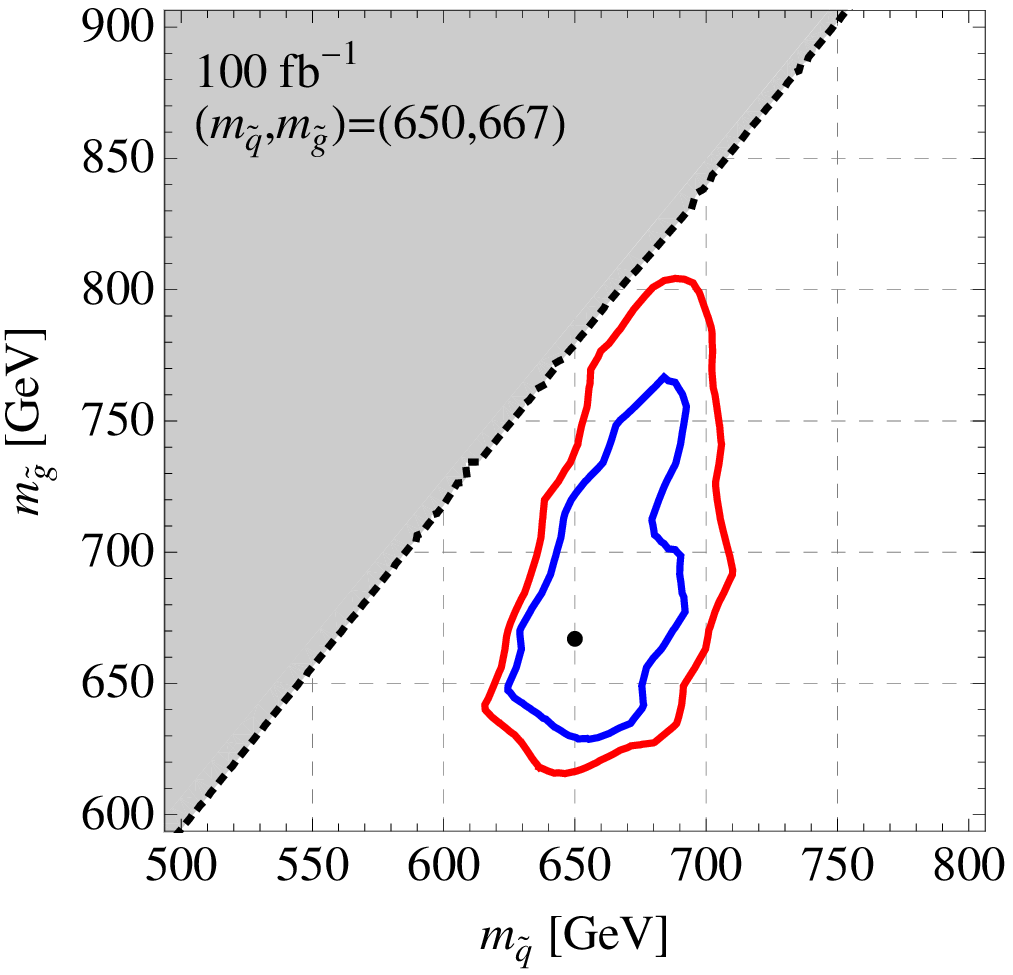}}
\hspace{2mm}
\subfigure[\ck{}]{\includegraphics[scale=0.72]{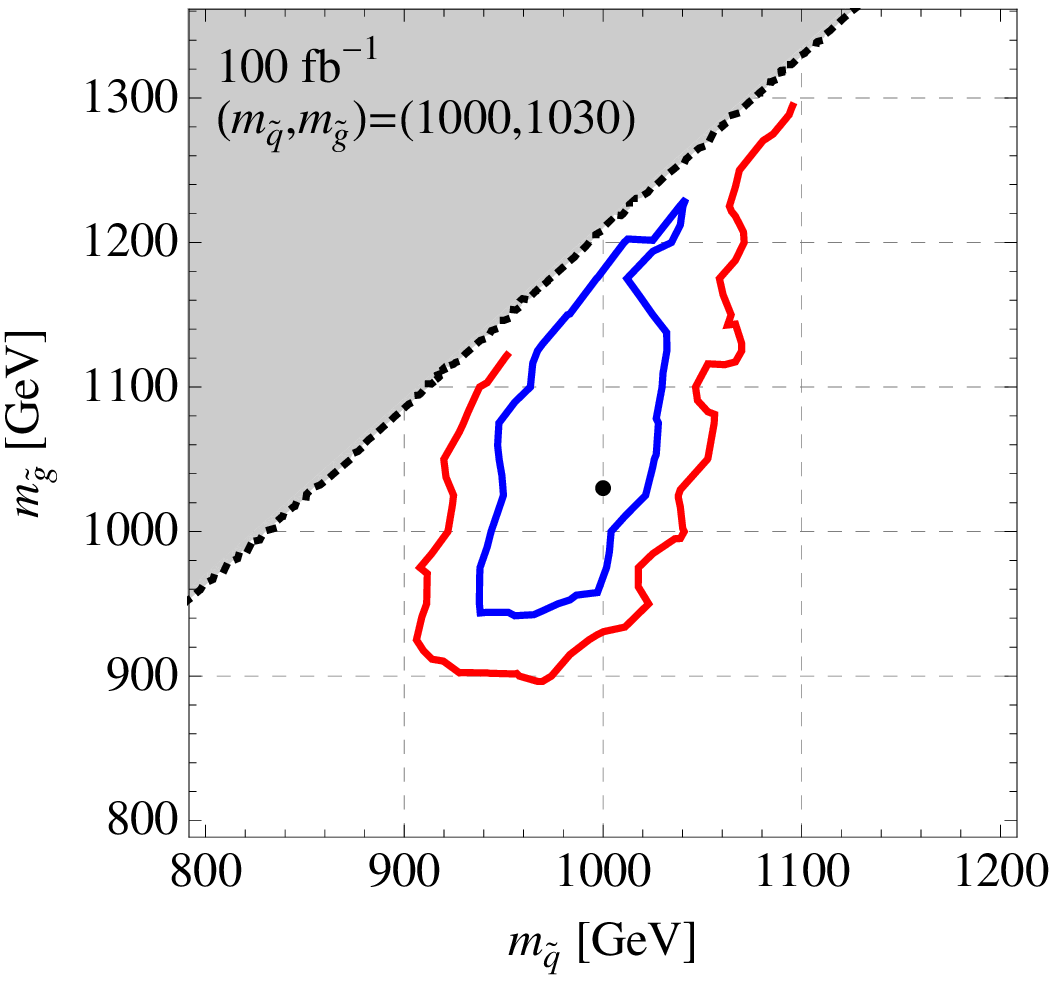}}
\end{center}
\caption{$2\sigma$ and $3\sigma$ contours obtained by fitting the shape
  of the SUSY signal
di-jet mass spectrum for ({\it a}) $\msq=650$~GeV, and ({\it b})
$\msq=1$ TeV. As in Fig.~\ref{fig:csec}, the shaded area on the
upper-left is excluded because the slepton is either lighter than
neutralino or below $100$ GeV. 
The jaggedness of the contours is due to Monte Carlo statistics.}
\label{fig:contour}
\end{figure}

Fig.~\ref{fig:contour} illustrates the results of our fit of the di-jet
mass spectrum after the cuts of Eq.~(\ref{eq:cuts}) for the two SUSY
cases.  The inner (outer) line corresponds to the $2\sigma$ ($3\sigma$)
contour where $\Delta\chi^2$ = 6.2 (11.8) above the minimum value of
$\chi^2$.  The reader may be surprised to see that the di-jet
distribution also serves to constrain the gluino mass. To understand
this, we note that starting from the best fit value, as we reduce
$m_{\tg}$ (keeping $m_{\tq}$ fixed), squark decays to gluino open up and
start to eat into the branching ratio $B(\tq \to q\tz_1)$. When the
gluino is only just a very tiny bit below $m_{\tq}$, it decays via $\tg
\to q \tq^*$, with the virtual squark almost on its mass shell, and the
additional quark jet being too soft for detection. The virtual
almost-on-shell-squark then decays via $\tq^* \to q\tz_1$ and the fit is
almost unaffected because events from squark decays to gluinos cannot be
kinematically distinguished from events with squarks decaying to the
LSP. However, as the gluino-squark mass difference increases, (1)~gluino
decays of squarks become more significant and begin eating into the
branching fraction for LSP decays, and (2)~the quark daughter of the
squark becomes harder and so more likely to be detected. Indeed, if all
$\tq \to \tg q$ decays would be vetoed by our jet-multiplicity cut, all
that would happen would be a loss of statistics, causing the ellipse to
widen, and ultimately open up, for lower values of $m_{\tg}$. However,
before this can happen, the three-body gluino decay still has
quasi-two-body kinematics, but with an off-shell squark lighter than
$m_{\tq}$. As a result the template gluino events will be similar to
squark events with a smaller squark mass; this alters the shape of the
corresponding $m_{jj}$ distribution, enabling the ellipse to close at
low gluino masses.

If on the other hand, starting from the best fit value we now go up in
$m_{\tg}$ (again keeping $m_{\tq}$ fixed), the gluino decays via $\tg
\to q\tq$. Initially, the quark from the gluino is very soft, and
gluinos simply act as a source of on-shell squarks. Thus $\tq\tg$
production just leads to better statistics, and so lead to a  more
precise determination of $m_{\tq}$, as we have already noted. As the
template value of $m_{\tg}$ is increased with $m_{\tq}$ fixed, the $E_T$
distribution of the daughter squark from gluino decay, and hence of its
daughter jet, is affected. Moreover, the quark daughter of the gluino
becomes increasingly more likely to be detected, and events with gluinos
are more likely to be vetoed.  These effects combine and together lead
to an increase of $\Delta\chi^2$. Note, however, that the quark daughter of
the gluino will not be significantly harder if the template squark mass
is also increased along with the template gluino mass, so that the
template mass gap is still small. In this case, the effect that we have
just mentioned is somewhat ameliorated, causing the error ellipses to
tilt to the right as well as narrow in the squark direction as they
extend to larger gluino (and squark) masses.\footnote{If however, $m_{\tg}\gg
m_{\tq}$ the rate of gluino events would be less frequent, and squark
events for this case would differ from those for the test point only by
changes in the squark $E_T$ distribution and the event rate due to the
much larger value of $m_{\tg}$. It is conceivable, therefore, that there
is another region (well into the grey region in the figure) where
$\Delta\chi^2$ is reduced. We have not investigated whether this does
indeed occur.}

 We see from Fig.~\ref{fig:contour} that this analysis constrains the
squark mass at $2\sigma$ to:
\be 
\begin{array}{ccl}
(\msq,\mg)=\mbox{(650,667) GeV} &:& \msq=635-690\mbox{~GeV}\\
(\msq,\mg)=\mbox{(1,1.03) TeV} &:& \msq=935-1040\mbox{~GeV}
\end{array}
\label{eq:dijetconstraint}
\ee
The better precision
in the first case is partly due to the fact that the higher statistics
allows us to make use of the dip structure that we discussed in
Sec.~\ref{sec:sqmass}.  We note also that the gluino mass is constrained
to lie between 630-760~GeV (940-1220~GeV) in the two cases, with the
true value being significantly closer to the lower end of the range.

\section{A comparison with squark mass determination using $m_{T2}$}

In this section we briefly examine how the extraction of the squark mass
from the di-jet mass distribution compares with the corresponding
determination using the $m_{T2}$ variable \cite{mt2} that
has received extensive attention during the last few years. There is no
question that, as we have already mentioned in Sec.~\ref{sec:intro} an
$m_{T2}$ analysis can potentially lead to a simultaneous determination
of gluino and LSP masses \cite{choi} if squarks are heavy. Here, we
focus our attention on the more difficult case that the squark is
essentially degenerate with the gluino, in which case the kink structure
on which the gluino and LSP mass determination is based is less evident
\cite{choi}.  For comparison purposes, we focus on acollinear di-jet
events with the set of cuts Eq.~(\ref{eq:cuts}) for the same test cases
in Fig.~\ref{fig:contour}. As in Sec.~\ref{sec:massanal}, we assume that
the SM backgrounds ($Z+jj$ and $W+jj$) can be subtracted, but still
include these in our analysis  of the statistical significance of the signal.

The $m_{T2}$ variable is an extension of the well-known transverse mass
($m_T$)~\cite{bargerphillips1987} for the case of two invisible final
state daughters.  For the one step cascade decay considered here
(see Eq.~(\ref{eq:cdec})), we have:

\[ m_{T2}(m_x) = \min_{\pT(D_1) + \pT(D_2) = {\bf \eslt} } [max(m_T^{(1)},m_T^{(2)})] \]
with
\[ m_T^{(i)} = \sqrt{m_x^2 + 2(E_T(q_i) E_T(D_i) - \pT(q_i).\pT(D_i))}\;, \]
where $D = \tz_1$, $P=\tq$, $m_x$ is the trial daughter mass and the final
state quark masses have been neglected.  The minimization on the
right-hand-side is carried out over the partitions of the ${\bf \eslt}$
vector. Since $m_{T2}(m_x = m_D) \leq m_P$, the value of
$m_{T2}^{max}(m_D)$ determines the parent's mass. However, because the
LSP mass ($m_D$) is not known, in general it is not possible to obtain
the parent mass $m_P$ (the squark mass in our case). It can be shown
that for {\it any} value of the trial LSP mass $m_x$, 
\begin{equation}
m_{T2}^{max}(m_x) = E_0 + \sqrt{E_0^2 + m_x^2} \label{eq:mt2max}\;,
\end{equation}
with $E_0 = (m_{\tq}^2 - m_{\tz_1}^2)/2m_{\tq}$. Thus for any value of
$m_x$, a determination of $m_{T2}^{max}$ serves to determine\footnote{Of
course, this simple analysis will be affected by the altered kinematics
from gluino events contributing to the di-jet sample, and also by QCD
radiative corrections.}  $E_0$. Assuming that $m_{\tz_1}^2/m_{\tq}^2 \ll
1$, this reduces to a determination of the squark mass. In this
respect, the information that we get from an $m_{T2}$ analysis is
identical\footnote{It has been suggested \cite{ben} that
if the squark pair is produced with a high transverse momentum ($\sim m_{\tq}$)
it may be possible, in principle, to determine both $m_{\tq}$ and $m_{\tz_1}$ separately
using the $m_{T2}$ procedure. We note, however, that under 1\% of squark pairs will be
produced with these large values of the pair transverse momentum, and so do not pursue
this any further in our study.} to that we obtain using the method that we have described in
Sec.~\ref{sec:sqmass}.  In our analysis we take $m_x=0$ to obtain,
\begin{equation}
m_{T2}^{max}(m_x=0) = 2 E_0 \approx m_{\tq}\;. \label{eq:mt2max2}
\end{equation}

\begin{figure}
\begin{center}
  \subfigure[]{\includegraphics[width=8.3cm]{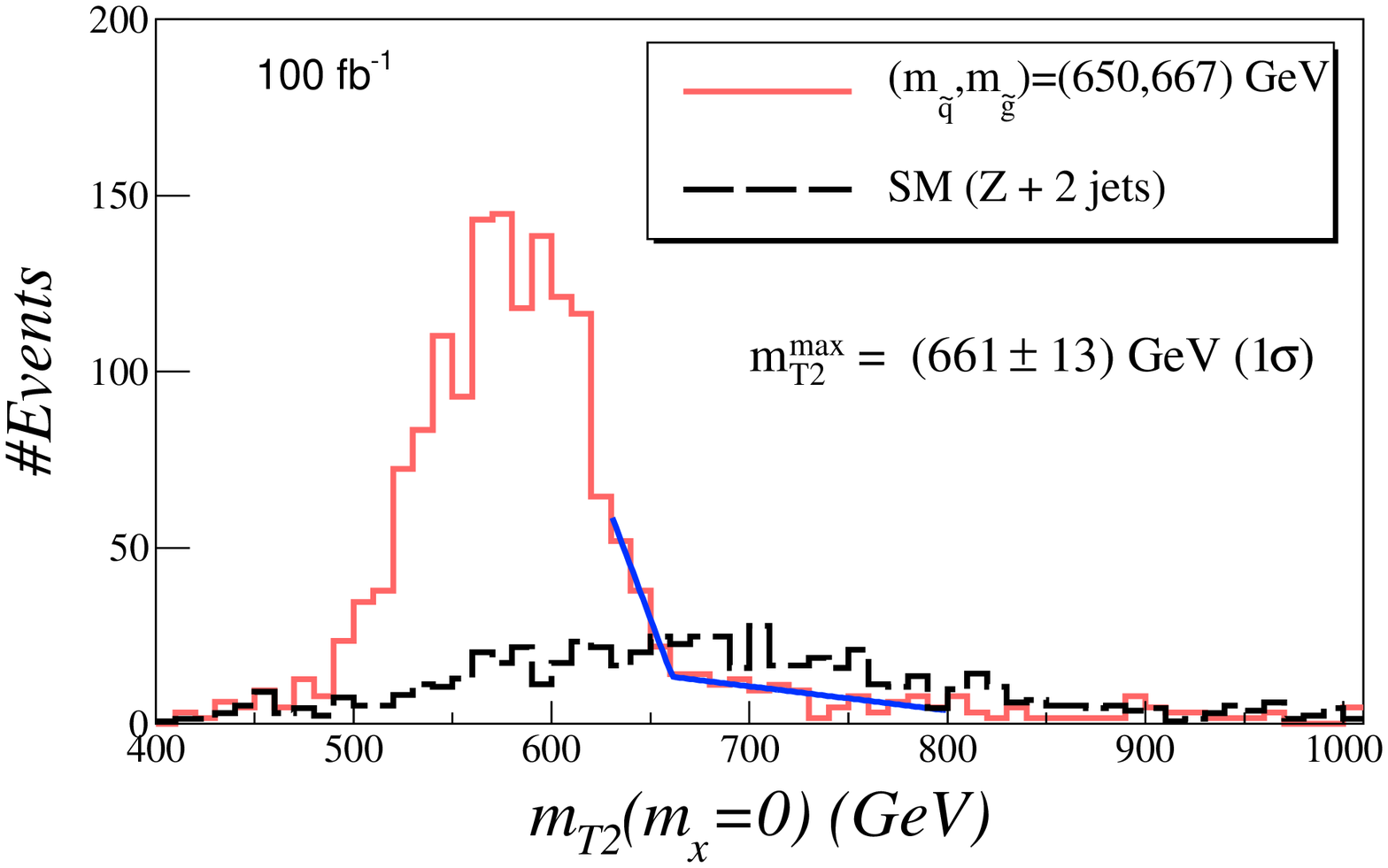}}
  \subfigure[]{\includegraphics[width=8.3cm]{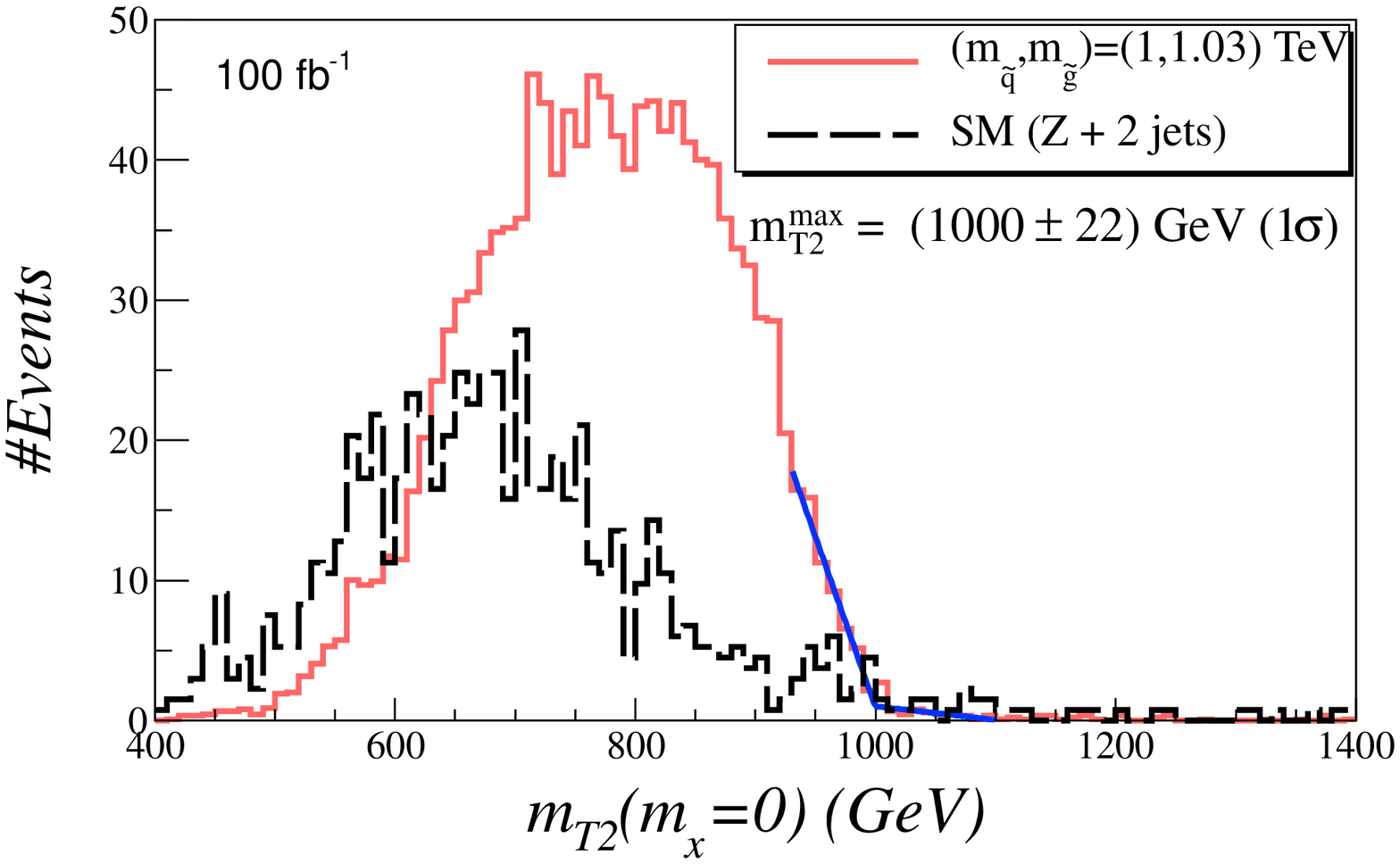}}\end{center}
\caption{The $m_{T2}$ distribution for ({\it
a})~$(m_{\tq},m_{\tg})=(650,667)$~GeV and ({\it b})~$(m_{\tq}, m_{\tg})=
(1000,1030)$~GeV SUY points in Fig~\ref{fig:contour} after the cuts in
Eq.~(\ref{eq:cuts}) of the text. The trial LSP mass is set to zero. The
straight lines show the ``linear kink fit'' to the signal, and is used
to extract $m_{T2}^{max}$. The dashed histogram shows the $m_{T2}$
distribution for the dominant SM BG (Z + 2 jets). We also show the
$m_{T2}^{max}$ value obtained from the fit and its $1\sigma$ error.}
\label{fig:mt2}
\end{figure}

We show the $m_{T2}(m_x = 0)$ distributions for the SUSY points
$(m_{\tq},m_{\tg})=(650,667)$~GeV and $(1000,1030)$~GeV after the cuts
of Eq.~(\ref{eq:cuts}) in Fig.~\ref{fig:mt2}. We see that the
distributions show distinct edges around $660$~GeV and 1 TeV for the
$m_{\tq} = 650$~GeV and 1 TeV cases, respectively. To extract the
$m_{T2}^{max}$ value from the $m_{T2}$ distributions we fit a ``linear
kink function'' for a selected bin range on either side of the visible
edge. The error for the fit includes the signal and BG statistical
errors added in quadrature, as discussed in Sec.\ref{sec:massanal}.
With this procedure, for the two case studies we obtain at $2\sigma$:
\be 
\begin{array}{ccl}
(\msq,\mg)=\mbox{(650,667) GeV} &:& \msq=640-682\mbox{~GeV}\\
(\msq,\mg)=\mbox{(1,1.03) TeV} &:& \msq=970-1030\mbox{~GeV}
\end{array}\label{eq:mt2fit}
\ee
We see that for the lighter squark mass case for which we were able to
use the dip structure, the error is essentially the same with both
methods, whereas for the heavier squark case, the $m_{T2}$ analysis
yields a somewhat lower error\footnote{We have verified that both the
central and error values for $m_{\tq}$ is weakly dependent on the choice
of the trial mass $m_x$. The magnitude of the error is, however,
somewhat sensitive to the number of bins on both side of the kink that
we use in our fit. We have shown a conservative range in
Eq.(\ref{eq:mt2fit}) above. }.

The squark mass result for the first SUSY point is shifted towards
higher values than the true value.  This is due to the $\tg\tq$
contamination, which tends to mimic $\tq\tq$ events with a heavier
squark mass. On the other
hand, the 1 TeV squark case has a much smaller contamination of $\tq\tg$
events ($< 10\%$) and $m_{T2}^{max}$ in this case provides the correct
value for $m_{\tq}$.

Before ending this section, we list the pros and cons of the extraction
of the squark mass from  $m_{T2}$  or from the di-jet mass distribution
described here. 
\begin{itemize}
\item Clearly, the di-jet mass method  is inapplicable if the squark is much
  heavier than the gluino, since then the di-jet signal from
  squark production is very small. In this regime, $m_{T2}$ is clearly
  superior as it potentially offers the possibility of determining both
  $m_{\tg}$ and $m_{\tz_1}$.

\item If $m_{\tq} \simeq m_{\tg}$, and the integrated luminosity of the
  LHC is large enough to enable us to utilize the dip structure in the
  $m_{jj}$ mass distribution, the two procedures yield the same
  information ({\it i.e.} a measure of $m_{\tq}$ if $m_{\tz_1}^2\ll
  m_{\tq}^2$), and with comparable precision. The di-jet mass
  distribution also offers the possibility of constraining
  $m_{\tg}$. This can then be compared with a direct determination of
  $m_{\tg}$ from the independent multi-jet event sample using, for
  instance, $m_{\rm eff}$. Consistency of the gluino mass obtained from
  these {\it independent} data samples would bolster the case that the
  new physics is indeed supersymmetry. 

\item In our analysis, we have assumed that the squarks that lead to
  di-jet events (mostly $\tu_R$ and $\td_R$ for mSUGRA) are essentially
  degenerate in mass. If these had a small but significant mass
  splitting, the shape of the $m_{jj}$ distribution would likely be
  distorted and we would extract some sort of averaged mass. The
  $m_{T2}$ analysis, on the other hand, relies on an end-point (which
  the mass splitting may smear) and would yield the larger of the squark
  masses. 

\item Our method for extracting $m_{\tq}$ relies only on the jet
  energies and angles, whereas $m_{T2}$ also requires $\eslt$. While the
  jet energy scale is likely to be the largest experimental systematic
  uncertainty for mass extraction from the $m_{jj}$ distribution, the
  $m_{T2}$ procedure relies on the entire detector,  and  so is likely
  to have 
  different experimental systematics. 

\end{itemize}

\section{Summary}

We have proposed a new way to determine the squark mass using templates
to fit the {\it shape} of the invariant mass distribution of acollinear
dijet events at the LHC. Standard model backgrounds are suppressed by
cuts on $\alpha,~\Delta\phi_{jj}$, $E_{T}^{tot}$ and veto of isolated
leptons. Our analysis is independent of many details of the underlying
model, and requires only that at least some types of squarks have a
large branching ratio for direct decays to the LSP. It yields a measure
of the squark mass if $m_{\tz_1}^2 \ll m_{\tq}^2$. While the signal may
have sizable $\eslt$, imposing a $\eslt$ cut does not lead to a
qualitative improvement on the precision with which the squark mass is
determined.

We emphasize that the squark mass is determined by a fit to only the
shape of the signal di-jet invariant mass spectrum. If $m_{\tg} \simeq
m_{\tq}$, the signal may also get a significant contribution from
$\tg\tq$ production which, like squark pair production, is fixed by SUSY
QCD\footnote{To be technically precise, we do assume that
intra-generational squark mixings are negligible.} in terms of just
$m_{\tq}$ and $m_{\tg}$. We do not make use of the magnitude of the
signal which would depend on model-dependent branching fractions for
squark and gluino decay. Indeed, the gluino contamination and/or the
dependence of the squark $E_T$ spectrum on $m_{\tg}$ also allows us to
constrain $m_{\tg}$, albeit with considerably lower precision than
$m_{\tq}$.   

As discussed in Sec.~\ref{sec:sqmass} and in more detail in the
Appendix, the cuts on $\alpha$ and $E_T^{tot}$ serve a dual role.  Aside
from background reduction, with appropriate choices $\alpha > 0.5$ (the
natural value to discriminate the signal from mismeasured QCD di-jet
events) and $E_T^{tot} \agt m_{\tq}$ cuts, we have shown that signal
events with intermediate values of $m_{jj}$ pass the cuts with reduced
efficiency compared to events with high and low di-jet masses. The
resulting dip structure in the di-jet mass distribution (see the inset
in Fig.~\ref{fig:scatter}{\it a}) is very sensitive to the value of
$m_{\tq}$, and so increases the precision with which the squark mass can
be extracted if the integrated luminosity is high enough to allow the
implementation of a hard
enough cut on $E_T^{tot}$.

We have performed two case studies, each with the gluino just slightly
heavier than the squark, to investigate the precision with which it is
possible to extract the squark mass from the dijet distribution. Our
results are shown in Fig.~\ref{fig:contour}.  We see that for the 
$m_{\tq}=650$~GeV  case for which the dip structure in the di-jet mass
distribution should be visible in LHC experiments with an integrated
luminosity of 100~fb$^{-1}$, the squark mass is constrained to lie
between 630-690~GeV at $2\sigma$, while for a 
1~TeV squark for which the  100~fb$^{-1}$ is not sufficient to utilize
the dip structure,  the corresponding range is 935-1040~GeV.   
We find that in the former case the precision that we obtain is 
essentially the same as that we would obtain using  $m_{T2}$, but the
systematics of the two measurements are quite different.

The method that we have proposed using the analysis of di-jet events
from supersymmetry to determine the squark mass can be generalized to
other beyond the SM theories and used to determine the characteristic
heavy particle mass. The invariant mass construction relies only on
kinematical cuts, the assumption of a low mass of the stable (invisible)
particle and a large enough branching ratio for the two body decay of the
heavy particle into a jet and the invisible lighter particle.

\section*{Acknowledgments} 

\noindent XT thanks the UW IceCube collaboration and the UW Phenomenology
Institute for making his visit to the University of Wisconsin, where
part of this work was carried out, possible. This research is supported
in part by grants from the US Dept. of Energy
and by the Fulbright Program and CAPES (Brazilian Federal Agency for
Post-Graduate Education).

\appendix
\section*{Appendix: The Dijet Invariant Mass}
\label{sec:app}

The di-jet invariant mass in Fig.~\ref{fig:scatter} shows
distinct features after the cuts in Eq.~(\ref{eq:cuts}).
Here we present a discussion of the kinematics
behind these distinct shapes and how they relate to the squark mass. For
tractability, we neglect QCD
radiation, SUSY contamination and SM backgrounds. 

As we will see below, the dip structure in the $m_{jj}$ distribution
that we have discussed at length in the text mostly arises from the
$E_T^{tot} \equiv E_T(j_1)+E_T(j_2)$ and $\alpha$ cuts, hence we shall
omit the $\Delta\phi_{jj}$ cut to start with.  Since $m_{jj}$,
$\alpha$ and $E_T^{tot}$ are invariant under longitudinal boosts along
the beam axis (the $Z$-direction) and
rotations, we can restrict our analysis to the squark-pair center of
mass frame, with the momentum of both squarks in the $XZ$ plane:
\begin{eqnarray}
\tq_1^{\mu} & = & \gamma m_{\tq}(1,\beta \sin\theta,0,\beta \cos\theta)\;, \nonumber \\
\tq_2^{\mu} & = & \gamma m_{\tq}(1,-\beta \sin\theta,0,-\beta \cos\theta)\;, \label{eq:sqmom}
\end{eqnarray}
where $\beta$ is the speed of the squarks, and
$\gamma=1/\sqrt{1-\beta^2}$. The jet energy in the squark rest frame is
$E_0 = (m_{\tq}^2 - m_{\tz_1}^2)/2m_{\tq}$. Its value in the 
squark-pair rest frame is changed by the boost, but is typically of this
order.  A hard jet (as required by $E_T^{tot} \gtrsim 2E_0$) results when the daughter
quark is roughly collinear with its parent squark.  In order to satisfy
the hard $E_T^{tot}$ cut, we will for simplicity, make the
approximation that the hardest quark is parallel to its parent squark,
which fixes its four-momentum to be,
\begin{equation}
q_1^{\mu} = \gamma E_0(1+\beta)(1,\sin\theta,0,\cos\theta).
\end{equation}
However, $\theta_{\tq_2 q_2}$, the angle between $\tq_2$ and $q_2$ in
the squark center-of-mass frame is not necessarily restricted to small
values and $q_2$ can be emitted in any direction, although, before cuts,
$\theta_{\tq_2 q_2} = 0$ still is the most likely value if the squark is
significantly boosted.  The four-momentum of the second quark can be
written as:
\begin{eqnarray*}
q_2^{\mu} & = &
E_0\left[\gamma(1+\beta\cos\theta_0),-\gamma\sin\theta(\cos\theta_0+\beta)
+ \cos\theta\sin\theta_0\cos\phi_0,\sin\theta_0\sin\phi_0, \right.\\ &&
\left.-\gamma\cos\theta(\cos\theta_0+\beta)
-\sin\theta\sin\theta_0\cos\phi_0\right]\;,
\end{eqnarray*}
where $\pi-\theta_0$ and $\phi_0$ are the polar and azimuthal angles of
$q_2$ in the $\tq_2$ rest frame (with axes oriented so that $\tq_2$ is
moving along the negative Z-axis), respectively. The $q_2$-$\tq_2$ angle
in the squark CM frame ($\theta_{\tq_2 q_2}$) is then related to
$\cos\theta_0$ through:
\[\cos\theta_{\tq_2 q_2}= \frac{\beta + \cos\theta_0}{1+\beta\cos\theta_0}, \],
while the dijet invariant mass ($m_{jj}$) is given by
\begin{equation}
m_{jj}  =  
\sqrt{2}\gamma E_0 (1+\beta)\sqrt{1+\cos\theta_0}\;.  \label{eqn:vars1}\\
\end{equation}

We now  proceed to investigate the shape of the $m_{jj}$ distribution under the
constraints:
\begin{equation}
\alpha > 1/2\;{\rm and }\; E_T^{tot} > 2 E_0\;. \label{eq:const}
\end{equation}
Toward this end, we consider three distinct cases:
\bi
\item A) $\cos \theta_{\tq_2 q_2} \approx 1 \Leftrightarrow \cos
  \theta_0 \approx 1$, corresponding to high $m_{jj}$ values,
\item B) $\cos \theta_{\tq_2 q_2} = 0 \Leftrightarrow \cos \theta_0 =
  -\beta,$ corresponding to intermediate $m_{jj}$ values, and 
\item C) $\cos \theta_{\tq_2 q_2} \approx -1 \Leftrightarrow \cos
  \theta_0 \approx -1,$  corresponding to low $m_{jj}$ values.
\ei

{\it Case} A). This configuration corresponds to $q_2$ emitted nearly
along its parent direction. Completely back to back jets ($\theta_0 =
0$) always have $\alpha \leq 1/2$ and therefore can never satisfy
Eq.(\ref{eq:const}). Thus, we take $\theta_0 = 0 +\epsilon$, where $0
<\epsilon \ll 1$. Then $\alpha$ and $E_{T}^{tot}$ become:
\begin{equation}
\alpha \approx \frac{\sin\theta}{2} - \epsilon\sqrt{\frac{1-\beta}{1+\beta}}\frac{\cos\theta\cos\phi_0}{2}\;\;{\rm and }\;\; E_T^{tot} \approx 2E_0[\sqrt{\frac{1+\beta}{1-\beta}}\sin\theta - \epsilon\frac{\cos\theta\cos\phi_0}{2}]  \nonumber
\end{equation}
We see from the expression for $\alpha$ above that the constraint
$\alpha > 0.5$ eliminates events with back-to-back (or nearly
back-to-back) jets, as does the $\Delta\phi<1.5$ cut in (\ref{eq:cuts})
of the text. This is, of course, why it so effectively reduces
the QCD background \cite{rts}. The highest di-jet mass events after the
cuts nevertheless arise from acollinear hard jets with large opening
angle between them, though without the $\alpha$ cut this distribution
would extend out to even higher $m_{jj}$ values.

{\it Case} B). This configuration corresponds to the case where $q_2$ is
emitted perpendicular to its parent direction and consequently
also perpendicular to $q_1$.
The expressions for $\alpha$ and $E_{T}^{tot}$ are:
\begin{equation}
\alpha = \sqrt{\frac{1-\beta}{2}}\sqrt{\cos^2\theta\cos^2\phi_0 + \sin^2\phi_0}\;\;{\rm and }\;\; E_T^{tot} = \sqrt{\frac{1+\beta}{1-\beta}}E_0[\sin\theta+(1-\beta)\sqrt{\cos^2\theta\cos^2\phi_0 + \sin^2\phi_0}] \label{eq:caseB}
\end{equation}
If we impose $\alpha > 1/2$, we have:
\[ \sqrt{\cos^2\theta\cos^2\phi_0 + \sin^2\phi_0} > \frac{1}{\sqrt{2(1-\beta)}} \Rightarrow \beta < 1/2 \]
For $\phi_0 = \pi$ or 0 we have:
\[ E_T^{tot} = \sqrt{\frac{1+\beta}{1-\beta}}E_0[\sin\theta+(1-\beta)|\cos\theta|] \]
The above expression is maximum for $\sin^2\theta = 1/(2-2\beta + \beta^2)$, which gives:
\[ E_T^{tot} < \sqrt{\frac{1+\beta}{1-\beta}}E_0 \sqrt{1+(1-\beta)^2} < \sqrt{\frac{15}{4}}E_0 \]
for $\beta < 1/2$. Therefore, for $\phi_0 = \pi$ or 0 ({\it i.e.} the
quarks, the squarks and the proton beam are all in one plane),
Eq.~(\ref{eq:const}) can never be satisfied. On the other hand, if we
look at extremely acoplanar configurations with $\phi_0 = \pi/2$ or
$3\pi/2$:
\[ E_T^{tot} = \sqrt{\frac{1+\beta}{1-\beta}}E_0[\sin\theta+(1-\beta)] \leq \sqrt{\frac{27}{4}}E_0 \approx 2.6 E_0\]
We see that such cases are allowed by our cuts.  However, imposing
$E_T^{tot} > 2E_0$ requires large values of $|\sin\phi_0|$ -- we have
verified numerically that, for Case B, $E_T^{tot} > 2E_0$ requires
$|\sin\phi_0| \gtrsim 0.75$ -- considerably restricting the phase space
for configurations that simultaneously satisfy the $\alpha$ and
$E_T^{tot}$ cuts.  This analysis, therefore, confirms the qualitative reasoning
in
Sec.~\ref{sec:sqmass} of the text for the existence of the dip at
intermediate $m_{jj}$ values.  More importantly this analysis
quantifies the magnitude
of the cut on  $E_T^{tot}$ that is needed for the appearance of the dip. 
Increasing the cut to $E_T^{tot} \gtrsim 2.6E_0$ would completely
suppress this configuration accentuating the dip even further, but this
harder cut would lead to a considerable reduction in the
statistics.  

{\it Case} C). Finally, in this case, $q_2$ is emitted anti-parallel to
its parent and consequently parallel to $q_1$. This is, of course, an
oversimplification, since the two would-be jets would then be merged
into a single jet by our cone algorithm for defining jets, which is why
we write the expression for $\alpha$ allowing an angle $\epsilon$
between $q_2$ and its parent squark Taking $\theta_0 = \pi -\epsilon$ ,
with $0 < \epsilon \ll 1$, we find that,
\begin{equation}
\alpha \approx \frac{\sin\theta (1-\beta)}{(1+\beta)\epsilon}\;\;{\rm and }\;\; E_T^{tot} \approx 2E_0\gamma \sin\theta\;, \label{eqn:casec} \nonumber
\end{equation}
independent of $\phi_0$. We readily see that this small $m_{jj}$ momentum
configuration can easily satisfy the 
cuts in Eq.~(\ref{eq:const}).  We point out that a {\it very large} squark
boost would make this configuration less likely. It is thus possible 
that a very hard cut on $E_T^{tot}$ would eliminate most events with
very low values of $m_{jj}$, causing the dip structure to wash out
altogether.

\begin{figure}
\begin{center}
\includegraphics[scale=0.72]{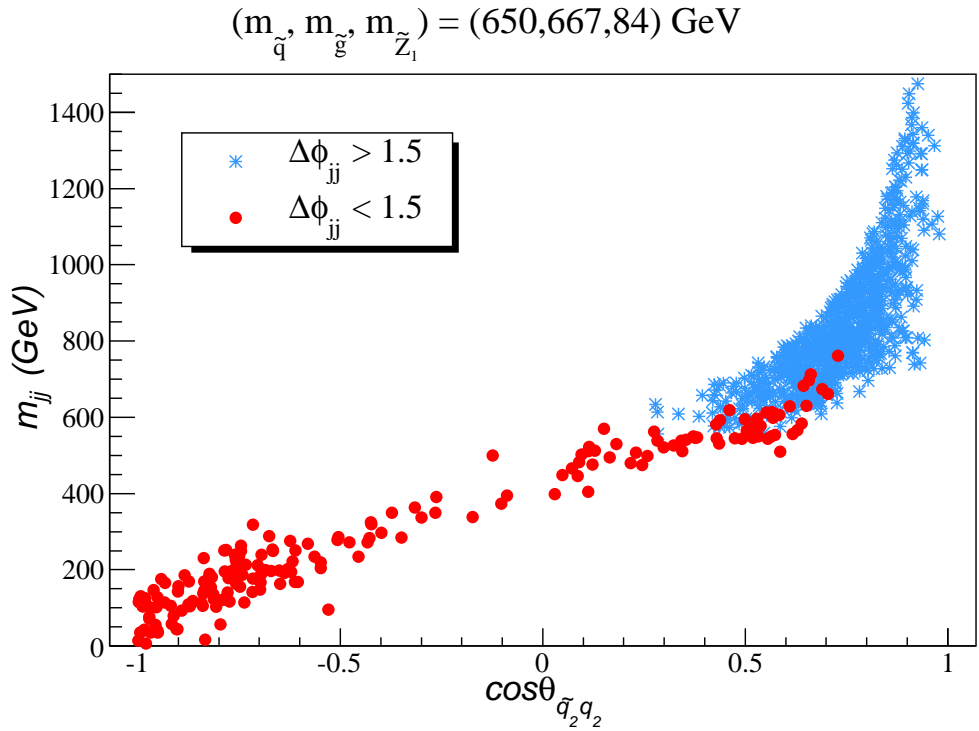}
\end{center}
\caption{A scatter plot of the invariant dijet mass ($m_{jj}$) versus
the quark-squark angle $\theta_{\tq_2 q_2}$ for
$(m_{\tq},m_{\tg},m_{\tz_1})=(650,667,84)$ GeV, after the $E_{T}^{tot} >
800$ GeV and $\alpha > 0.5$ cuts have been applied. The (blue) star points
have $\Delta\phi_{jj} > 1.5$, while the (red) dot points have
$\Delta\phi_{jj} < 1.5$}
\label{fig:cosmjj}
\end{figure}

To see how the semi-quantitative analysis with the three cases that we
have just discussed stands up to inclusion of {\it all} events from
squark pair production we show in Fig.~\ref{fig:cosmjj} a scatter plot of
$m_{jj}$ versus $\cos\theta_{\tq_2 q_2}$ in the squark CM frame, after
the $\alpha > 1/2$ and $E_T^{tot} > 800$~GeV cuts have been applied, but
with $\theta_{\tq_1 q_1}$ and $\phi_0$ allowed to take all values.  Events with an
azimuthal angle separation $\Delta\phi$ larger (smaller) than 1.5 are
depicted by stars (dots).  The points cluster at high $m_{jj}$ values
($\cos\theta_{\tq_2 q_2} \sim 1$), as expected, since the squark boost
causes the $\cos\theta_{\tq_2 q_2}$ distribution to peak at
$\theta_{\tq_2 q_2} =0$. We see that the bulk of these events have
$\Delta\phi > 1.5$.  On the other hand, for intermediate $m_{jj}$ values
($\cos\theta_{\tq_2 q_2} \approx 0$), which corresponds to Case B
discussed above, relatively few events pass the cuts.  Once we move to
large negative values of $\cos\theta_{\tq_2 q_2}$ (Case C), the
kinematical configuration once again satisfies the cuts, even though the
squark boost tends to suppress emission in this direction.  We thus see
how the double-peaked $m_{jj}$ profile shown in
Fig.~\ref{fig:scatter}{\it a} arises, once we require $E_T^{tot} >
700$~GeV, $\alpha>0.5$ and $\Delta\phi_{jj} < 1.5$ to increase the
signal cross-section and reduce the SM background. As
can be seen from Fig.\ref{fig:cosmjj}, the $\Delta\phi_{jj} < 1.5$ cut
suppresses the high $m_{jj}$ values but does not change the double-peaked
structure.

Finally, our analysis also makes clear that if we reduce the cut on
$E_T^{tot}$ and allow events with $E_T^{tot} \alt 2E_0$, kinematic
configurations of Case B will  be
allowed for any $\phi_0$ value and the cut suppression of
intermediate $m_{jj}$ values will no longer occur, causing the dip to
disappear. We then expect the
invariant mass distribution to present a single peak at intermediate
$m_{jj}$. This is indeed what happens in Fig.\ref{fig:scatter}{\it b} for the
$(m_{\tq},m_{\tg}) = (1000,1030)$ GeV point, where $E_0 = 490$ GeV and
the cuts are $E_{T}^{tot} > 700$ GeV, $\alpha > 0.5$ and
$\Delta\phi_{jj} < 1.5$.

%%%%%%%%%%%%%%%%%%%%%%%%%%%%%%%%%%%%%%%%%%%%%%%%%%%%%%

\end{document}